\documentstyle[sprocl,psfig]{article}

\newcommand{\g}{$\gamma$}
\newcommand{\mg}{$^{26}$Mg}
\newcommand{\mgng}{$^{26}$Mg(n,$\gamma$)$^{27}$Mg}

\title{
	ASTROPHYSICALLY RELEVANT NEUTRON CAPTURE NEAR THE BORDER
	OF STABILITY
}

\author{P.~Mohr and H.~Oberhummer}

\address{
	Institut f\"ur Kernphysik, Technische Universit\"at Wien,\\
	Wiedner Hauptstra{\ss}e 8--10, A--1040 Vienna, Austria
}

\author{H.~Beer}
\address{
	Forschungszentrum Karlsruhe, Institut f\"ur Kernphysik III,\\
	P.O.~Box 3640, D--76021 Karlsruhe, Germany
}

\date{}

\begin{document}

\maketitle

\abstracts{
The neutron capture cross section on \mg\ was measured in the
astrophysically relevant energy region from 25 keV to 220 keV. 
The experimental results
agree well with a calculation using the Direct Capture (DC) model
together with systematic folding potentials. This experiment
confirms for the \mgng\ reaction that the DC process is 
at least comparable to the Compound--Nucleus (CN) process. 
The reliability of
DC calculations is discussed, and we present some ideas for future experiments
which could reduce the theoretical uncertainties of DC calculations.
}

\section{Introduction}
\label{sec:intro}
Low--energy neutron capture is very important for the nucleosynthesis
in stars. Capture cross sections on neutron--rich stable nuclei
close to the border of stability at thermonuclear energies
($kT \approx 8 - 100~{\rm{keV}}$)
are required for $s$--process calculations. For the $\alpha$--rich freeze--out
in Type II supernovae and for the $r$--process
one needs capture cross sections on stable and unstable neutron--rich nuclei
at somewhat higher energies ($kT \approx 50 - 250~{\rm{keV}}$). 
The cross sections for the $s$--process
can be measured, but neutron capture experiments on unstable
nuclei are in general not possible.

Up to now calculations of $r$-process capture cross sections on unstable
nuclei have mainly been performed using the statistical Hauser-Feshbach
(HF) model where the capture cross section is derived from penetration 
probabilities and the level density
in the compound nucleus (CN); even though the importance of the Direct
Capture (DC) process was recognized in the last years it was not used
in $r$-process calculations. (See, e.g.\cite{gor96}: 
``Of course, in these cases, the validity
of the adopted Hauser-Feshbach rate predictions has to be questioned, the
neutron capture process being possibly dominated by direct electromagnetic
transitions to a bound final state rather than through a compound nucleus
intermediary\ldots This complication is neglected here.'')

\begin{figure}[ht]
\begin{center}
        \leavevmode
        \psfig{figure=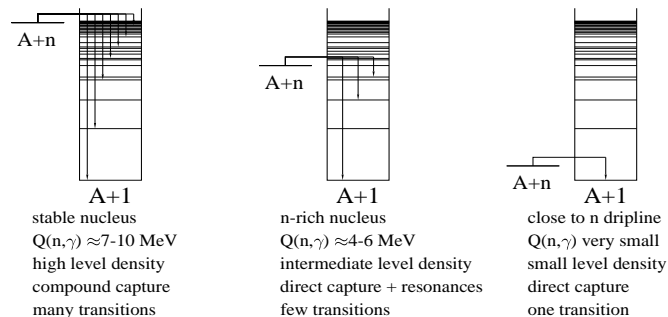,height=3.2cm}
\end{center}
\caption{
        \label{fig:level}
	Schematic presentation of energy levels relevant
	for (n,\g ) capture reactions on nuclei with high, intermediate, and
	low neutron separation energy (from left to right).
}
\end{figure}

A high level density in the CN is necessary for the application of the HF 
model. Almost all intermediate and heavy stable
nuclei fulfill this condition,
because the neutron binding energy is of the order of about 7--10 MeV.
However, for neutron-rich nuclei involved in the $r$-process the binding
energies decrease down to zero at the neutron dripline, and therefore
the level density in the CN becomes very small. This behavior is shown
schematically in Fig.~\ref{fig:level}.
The DC process can be analyzed experimentally for neutron-rich light
and some intermediate nuclei (especially with magic neutron number)
where the binding energies are of the order of 4--6 MeV. A series
of experiments was performed at the Karlsruhe Van de Graaff
accelerator (FZK):
$^{15}$N \cite{n15},
$^{18}$O \cite{o18},
$^{36}$S \cite{s36},
$^{48}$Ca \cite{ca48}.

These measurements have to be supplemented by DC calculations of
(n,\g) capture cross sections on more
neutron-rich unstable target nuclei in the isotopic chain. Experimental
information to improve these calculations can be obtained from (d,p) reactions
using accelerated neutron-rich fission fragments and inverse kinematics.

\section{Present Experiments}
\label{sec:pres}
\begin{figure}[ht]
\begin{center}
	\leavevmode
	\psfig{figure=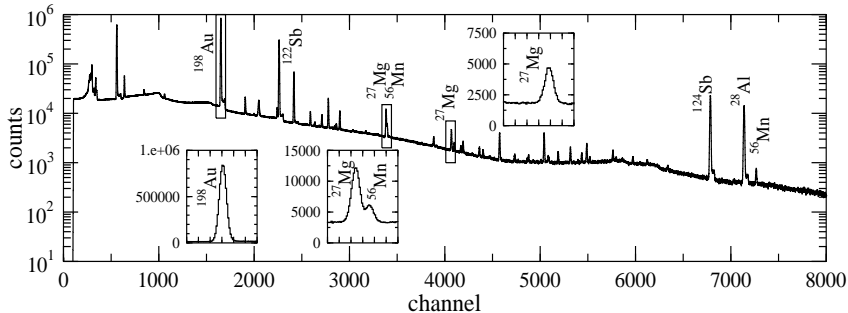,height=3.2cm}
\end{center}
\caption{
	\label{fig:spek_mg26}
	Typical $\gamma$--ray spectrum from the activation measurement of
	\mgng . The insets show the three regions of interest
	($^{198}$Au, $E_{\gamma} = 412$~keV; $^{27}$Mg,
	$E_{\gamma} = $~844, 1014 keV).
}
\end{figure}

For the measurement of small capture cross sections often the
activation method is used. This method is only applicable for unstable
residual nuclei with halflives longer than about 1 h. For nuclei with 
shorter halflives the fast cyclic activation method was developed at
the Van-de-Graaff laboratory of the Forschungszentrum Karlsruhe \cite{beer}: 
The sample is 
irradiated for a short time (comparable or shorter than the halflife),
then the sample is moved from the irradiation position to the counting
position, and the \g -rays following the $\beta$-decay of the residual
nuclei are detected using a high-purity germanium (HPGe) detector.
To gain statistics, this procedure can be repeated many times.

The neutrons for the activation experiment are generated by the
$^7$Li(p,n)$^7$Be reaction using beam currents of about 100 $\mu$A. 
At a proton energy close above the reaction threshold 
($E_{\rm p}^{\rm thres} = 1881~{\rm{keV}}$)
a quasi-thermal spectrum with $kT \approx 25~{\rm{keV}}$ is obtained.
At somewhat higher energies neutron spectra with a width
of about 20--30 keV were generated.

For the $^{26}$Mg experiment the sample consisted of a small tablet
of 98.79\% enriched $^{26}$MgO (diameter 6 mm, weight 90 mg) which was
sandwiched between two thin gold foils to determine the \mgng\ 
relative to the well-known $^{197}$Au(n,\g )$^{198}$Au cross section.
The residual nucleus $^{27}$Mg decays with a halflife of $9.458 \pm 0.012$ min
to $^{27}$Al which deexcites by \g -ray emission 
with $E_\gamma = 844~{\rm{keV}}$ (branching $71.8 \pm 0.4\,$\%) 
and $E_\gamma = 1014~{\rm{keV}}$ ($28.0 \pm 0.4\,$\%) \cite{endt90}.
A typical spectrum is shown in Fig.~\ref{fig:spek_mg26},
and the experimental cross section is compared to a DC calculation in
Fig.~\ref{fig:all}.

\section{Theoretical Analysis}
\label{sec:theo}
The capture cross section was calculated using the DC model \cite{kim81,kim87}:
\begin{equation} 
\sigma_{DC} \sim \sum_i C^2 S_i \cdot | < u_{NL,i}(r) \, | \,
{\cal O}^{\rm E{\cal{L}}/M{\cal{L}}} \, | \, \chi_{L} (r) > |^2
\label{eq:DC}
\end{equation}
with $C^2 S_i$: spectroscopic factors of the final states, 
$u_{NL,i}(r)$: bound state wave functions, $\chi_L(r)$: scattering
wave function, and $O^{E{\cal{L}}/M{\cal{L}}}$: 
transition operators of electromagnetic 
$E{\cal{L}}$ and $M{\cal{L}}$ transitions.

For the calculation of the wave functions we used 
systematic folding potentials
\begin{equation}
V_F(r) = 
        \lambda \cdot 
	\int \int \rho_P(r_P) \, \rho_T(r_T) \, v_{eff}(s,\rho,E) \; 
		d^3r_P \; d^3r_T
\label{eq:DF}
\end{equation}
with $\rho_P$ and $\rho_T$ being the densities of projectile and target,
and an effective
nucleon-nucleon interaction $v_{eff}$ \cite{devries87,kobos84,abele93}.
Compared to standard Woods-Saxon potentials the number of parameters
is reduced because the shape of the potential is determined by the
folding procedure.

The potential strength parameter $\lambda$ is adjusted to the binding energy
(bound states) and to the scattering length (scattering wave function)
leading to $\lambda$ values close to unity. The $p$-wave potential
has to be reduced by about 15\% in strength to obtain a good description
of the measured \mgng\ data. The spectroscopic factors were taken from
a $^{26}$Mg(d,p)$^{27}$Mg experiment \cite{meur74}.

\begin{figure}[ht]
\begin{minipage}{12.0cm}
\begin{minipage}[t]{7.9cm}
	\psfig{figure=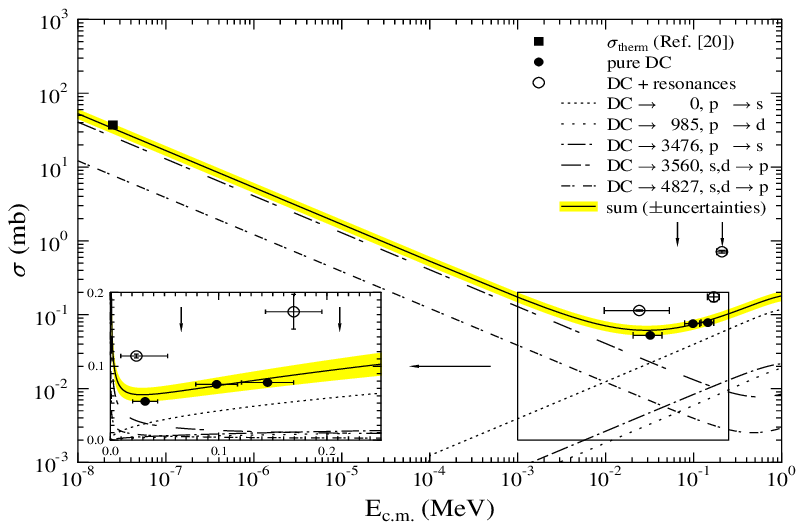,height=4.7cm}
\caption{
	\label{fig:all}
	Preliminary cross section of \mgng . The arrows mark known resonances
	\protect\cite{weig76}, and the experimental data points shown
	with open circles are affected by these resonances. 
	(Note that the horizontal error bars show only the FWHM of
	the neutron energy!)
	Additionally,
	the thermal capture cross section \protect\cite{mugh} 
	(full square) is well reproduced.
}
\end{minipage}\hfill
\begin{minipage}[t]{3.6cm}
	\psfig{figure=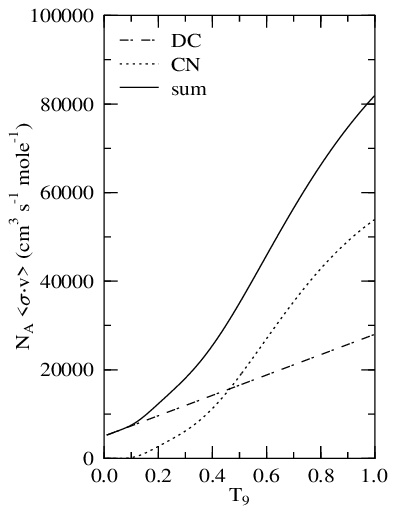,height=4.7cm}
\caption{
	\label{fig:rate}
	Reaction rate factor $N_A < \sigma \cdot v>$ of
	the neutron capture reaction \mgng .
}
\end{minipage}\hfill
\end{minipage}
\end{figure}

The results of the DC calculations for all strong transitions
of the capture reaction \mgng\ are shown in Fig.~\ref{fig:all},
and the astrophysically relevant reaction rate per particle pair 
$N_A < \sigma \cdot v >$ is shown in Fig.~\ref{fig:rate}.

\begin{table}[ht]
\begin{center}

\begin{tabular}{|c|c|c|c|c|}
\hline
$E_x ({\rm{keV}})$	& $J^{\pi}$	
	& $C^2 S_{\rm{exp}}$	& $C^2 S_{\rm{th}}$
	& $C^2 S_{(n,\gamma)}$	\\
\hline
0	& $1/2^+$	& $0.57 - 1.07$	& $0.43 - 0.70$	& $-$		\\
984.7	& $3/2^+$	& $0.37 - 0.80$	& $0.28 - 0.45$	& $-$		\\
1698.0	& $5/2^+$	& $0.13 - 0.31$	& $0.02 - 0.14$	& $-$		\\
3559.5	& $3/2^-$	& $0.40 - 0.56$	& $-$		& $0.37$	\\
3760.4	& $7/2^-$	& $0.55 - 0.80$	& $-$		& $-$		\\
4149.8	& $(5/2^+)$	& $0.03^*$	& $0.001 - 0.06$& $-$		\\
4827.3	& $(1/2^-)$	& $0.32^*$	& $-$		& $0.19^*$	\\
\hline
\multicolumn{5}{l}{\small{$^*$ assuming $(J^\pi)$ as given in column 2}}
\end{tabular}
\end{center}
\caption{
        \label{tab:spec}
	Spectroscopic factors of $^{27}$Mg = $^{26}$Mg $\otimes$ n
	taken from different experiments
	\protect\cite{meur74,turk88,pasch75,sinc76}
	and from shell-model calculations
	\protect\cite{meur74,moroz86,ben84}.
	Note that the spectroscopic factors of the states at $E_x =
	4149.8~{\rm{keV}}$ and 
	$4827.3~{\rm{keV}}$ were determined only in one experiment.
}
\end{table}

For stable nuclei the main uncertainties
are given by the spectroscopic factors.
An overview of spectroscopic factors obtained from different transfer reactions
\cite{meur74,turk88,pasch75,sinc76} 
and from shell-model calculations \cite{meur74,moroz86,ben84} is given in
Tab.~\ref{tab:spec}. Additionally, spectroscopic factors can be derived
very accurately
\cite{ca48}
from the thermal capture cross section by the ratio of 
experimental to calculated cross section 
(see Tab.~\ref{tab:spec}, last column).

The uncertainties in the potential are relatively small because the
$s$-wave potential which is the only relevant partial wave
at thermal energies can be properly adjusted, and the $p$-wave potential
should not differ by more than about 20\% from the $s$-wave potential.

For unstable nuclei far from stability the main uncertainties are
obviously given by the unknown spins, parities $J^{\pi}$ and
excitation energies $E_x$ of the relevant final states in the
residual nuclei, because E1 transition probabilities are proportional
to $E^3$ and E2 transition probabilities even scale with $E^5$.
Of course, the spectroscopic factors are also uncertain by at least a
factor of two
for these nuclei but these uncertainties are relatively small compared to the
$E^3$ and $E^5$ energy dependence of the transitions.

\section{Future Experiments}
\label{sec:futu}

Using accelerated neutron-rich fission fragments the input parameters of
DC calculations could be determined experimentally by (d,p) reactions
at energies in the order of 5--10 MeV/u. The relevant quantities 
which are spins and parities, excitation energies, 
and spectroscopic factors can
be derived from well-established DWBA analyses of the angular distributions.

Because of the inverse kinematics a relatively low background
can be expected in the proton spectra at laboratory backward angles
which correspond to forward angles in the center-of-mass system.
The deuteron elastic scattering can be measured at the same experiment
by detecting the recoil deuterons at laboratory forward angles
to derive the
deuteron-optical potential. The proton-optical potential for the
exit channel can be taken from potential systematics.
A sufficient count rate is expected if beam currents in the order
of ppA to pnA are obtained, of course, depending on the size of the
(d,p) cross section (about 1--100 $\mu$b).

\section{Conclusions}
\label{sec:conc}
We measured the DC contribution of the reaction \mgng\ using the
fast cyclic activation technique. The experimental results agree
reasonably well with our DC calculation. The importance of the
DC mechanism even for stable neutron-rich nuclei is again confirmed.

DC calculations of (n,\g ) reactions on unstable nuclei depend
sensitively on the input parameters., i.e., Q-value of the (n,\g )-reaction,
and $J^{\pi}$, $E_x$, and $C^2 S$
of the final states in the residual nuclei. These parameters can be
determined experimentally from (d,p) transfer reactions using 
accelerated neutron-rich fission fragments in reverse kinematics.

\section*{Acknowledgments}
We would like to thank the technicians G.~Rupp, E.~Roller, E.-P.~Knaetsch,
and W.~Seith from the Van-de-Graaff laboratory at the 
Forschungszentrum Karlsruhe.
This work is supported by Deutsche Forschungsgemeinschaft (DFG Mo739) and
Fonds zur F\"orderung der wissenschaftlichen Forschung (FWF S7307--AST).

\section*{References}

\end{document}